\documentclass[11pt,twocolumn,twoside]{article}
\usepackage{fully3d}
\usepackage{amssymb}

\usepackage{tabularray}
\usepackage{algorithm} % 导入算法环境宏包
\usepackage{algpseudocode} % 导入算法伪代码宏包

\begin{document}

%-------------------------------------------------------------------------------------------
%%%%% add your title here %%%%%
\title{PoCGM: Poisson-Conditioned Generative Model for Sparse-View CT Reconstruction} 

%%%%% add authors and affiliations here %%%%%
\author[1]{Changsheng Fang}
\author[1]{Yongtong Liu}
\author[1]{Bahareh Morovati}
\author[1]{Shuo Han}
\author[1]{Li Zhou}
\author[1,*]{Hengyong Yu}

\affil[1]{Department of Electrical and Computer Engineering,
           University of Massachusetts Lowel, Lowell, MA, US, 01854}
\affil[*]{Corresponding author, email: hengyong-yu@ieee.org }

% \affil[2]{Department of Radiology,
%           Recon University, City, Country}

%%%%% don't change these 2 lines %%%%%
\maketitle
\thispagestyle{fancy}

%-------------------------------------------------------------------------------------------
%%%%% add your summary (abstract) here               %%%%%%
%%%%% use footnotesize for this section              %%%%%%
%%%%% please stick to the customabstract environment %%%%%% 

\begin{customabstract}
In computed tomography (CT), reducing the number of projection views is an effective strategy to lower radiation exposure and/or improve temporal resolution. However, this often results in severe aliasing artifacts and loss of structural details in reconstructed images, posing significant challenges for clinical applications. Inspired by the success of the Poisson Flow Generative Model (PFGM++) in natural image generation, we propose a PoCGM (Poisson-Conditioned Generative Model) to address the challenges of sparse-view CT reconstruction. Since PFGM++ was originally designed for unconditional generation, it lacks direct applicability to medical imaging tasks that require integrating conditional inputs. To overcome this limitation, the PoCGM reformulates PFGM++ into a conditional generative framework by incorporating sparse-view data as guidance during both training and sampling phases. By modeling the posterior distribution of full-view reconstructions conditioned on sparse observations, PoCGM effectively suppresses artifacts while preserving fine structural details. Qualitative and quantitative evaluations demonstrate that PoCGM outperforms the baselines, achieving improved artifact suppression, enhanced detail preservation, and reliable performance in dose-sensitive and time-critical imaging scenarios.

%\bigskip

\end{customabstract}

%-------------------------------------------------------------------------------------------
%%%%% main text                                                %%%%%    
%%%%% remove the dummy content and put your own content here   %%%%% 
%%%%% feel free to choose your own section titles              %%%%% 
%%%%% you don't need to put the content in a separate tex file %%%%%

% dummy_content.tex shows how to add sections, figures, tables, formulas, and references
% remove the following line, it just adds dummy content
\section{Introduction}
Computed tomography (CT) is indispensable in medical imaging, providing detailed cross-sectional views for diagnosis and treatment. However, the conventional CT \cite{hsieh2003computed} typically requires a large number of projections to achieve high-quality images, resulting in increased radiation exposure. In scenarios like a C-arm machine, it is critical to reduce radiation dose or improve temporal resolution. Consequently, sparse-view CT \cite{singh2011adaptive}, which acquires fewer projection views, has become a promising strategy to reduce exposure and accelerate scans, yet it often suffers from streak artifacts and compromised image quality.

Reconstructing images from sparse-view data is an ill-posed inverse problem, traditionally addressed by analytical methods 
%(\emph{e.g.}, FBP)
or iterative algorithms with regularization. Recently, diffusion-based generative models, such as DDPM \cite{ho2020denoising}, DDIM \cite{song2020denoising}, and score-based approaches \cite{song2020score}, have shown promise for such tasks. EDM (Elucidating the Design Space of Diffusion-Based Generative Models) \cite{karras2022elucidating} systematically improves diffusion models by refining components like noise scheduling and sampling. Building on this, PFGM++ \cite{xu2023pfgm++} leverages electrostatics, treating \(N\)-dimensional data as charges in an augmented \(N+D\)-dimensional space. By tracking electric field lines with ODEs, PFGM++ enables efficient sampling from prior to target distributions. Notably, PFGM++ converges to EDM under certain conditions, showing that diffusion models are a special case of PFGM++. Furthermore, PFGM++ benefits from EDM's training and sampling algorithms with minor adjustments to prior noise and parameters.

% Building on these advances, PFGM++ \cite{xu2023pfgm++}models \textit{N}-dimensional data in an augmented \textit{(N+D)}-dimensional space via deterministic ODEs, offering faster and more stable sampling than SDE-based methods. When certain dimensional and radius parameters are chosen, PFGM++ converges to EDM, highlighting that widely used diffusion models are special cases within this framework. By tuning prior noise and a few variables, PFGM++ can adopt EDM’s training and sampling algorithms, unifying ODE efficiency with diffusion flexibility.

In this work, we propose PoCGM (Poisson-Conditioned Generative Model), an extension of PFGM++ tailored for sparse-view CT reconstruction. Leveraging its powerful prior modeling capabilities, PoCGM effectively suppresses artifacts and preserves high-fidelity images, broadening the applicability of sparse-view CT scanning for dose-sensitive and time-critical imaging scenarios. Our primary contributions are as threefold.
(1). First Application of PFGM++ to Sparse-View CT: To our knowledge, this is the first attempt to adapt PFGM++ for sparse-view CT reconstruction. By leveraging Poisson flow’s robust gradient guidance and efficient ODE sampling, PoCGM effectively reduces streak artifacts and noise, achieving high reconstruction fidelity even under extreme sparsity conditions. (2). Reformulation into a Conditional Generative Model: We reformulate PFGM++ into a conditional generative model by introducing paired data and conditional guidance during both of the training and sampling phases. This reformulation enables PoCGM to model the posterior distribution of full-view reconstructions conditioned on sparse observations, seamlessly integrating learned priors with observed measurements during the reverse Poisson flow to improve reconstruction accuracy. (3). Improved Performance over Baselines: Comprehensive evaluations show that PoCGM outperforms baseline reconstruction approaches in both quantitative metrics and visual quality. These results highlight PoCGM’s ability to suppress artifacts, recover structural details, and provide a practical solution for sparse-view CT reconstruction.
%%%%%%%%%%%%%%%%%%%%%%%%%%%%%%%%%%%%%%%%%%%%%%%%%%%%%%%%%%%%%%%%%%%%%%%%%%%%%%%%

\section{Methods}

\subsection{Problem definition}
\begin{figure}
    \centering
    \includegraphics[width=1\linewidth]{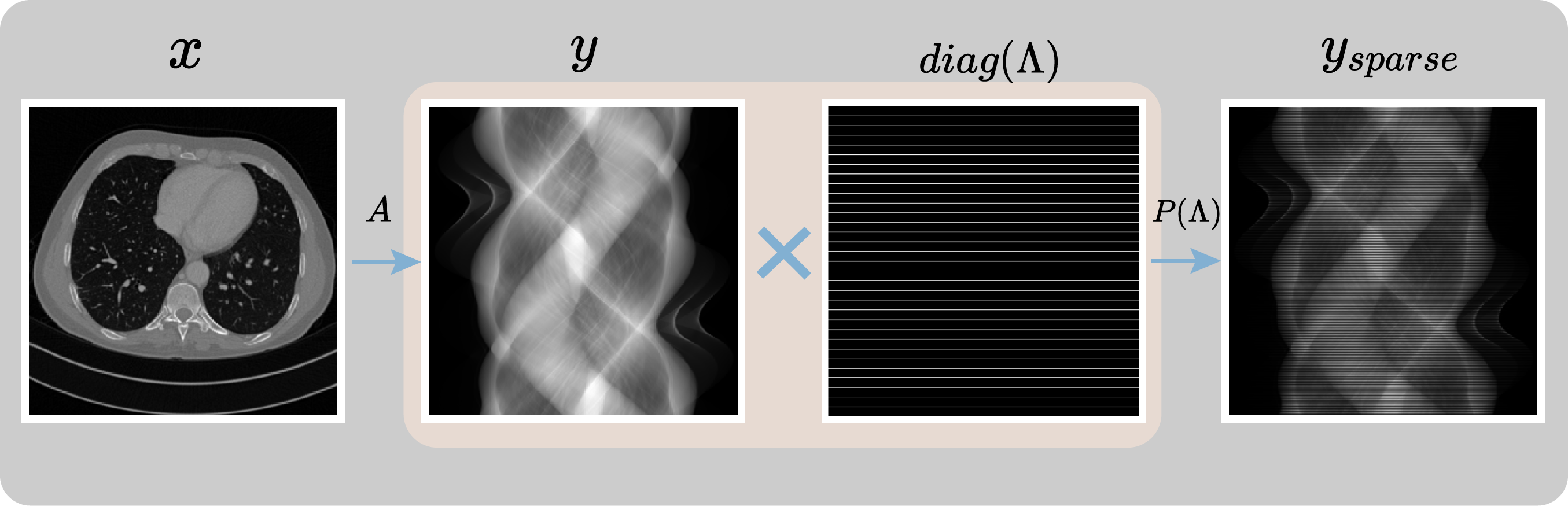}
    \caption{Data acquisition process of sparse-view CT}
    \label{fig:y=ax}
\end{figure}

Sparse-view data acquisition can be modeled as a sampling process from full projection data (see Fig. \ref{fig:y=ax}), where the sparse-view projection data \(\mathbf{y}_{sparse}\) is given by:
\[
\mathbf{y}_{sparse} = P(\Lambda)\mathbf{A}\mathbf{x},
\]
with \(\mathbf{A}\mathbf{x}\) representing full-view projection data, \(\mathbf{A}\) as the measurement operator (e.g., Radon transform), and \(P(\Lambda)\) as the sampling operator defined by mask \(\Lambda\). Directly applying traditional reconstruction methods, such as filtered backprojection (FBP), often produces sparse-view images \(\mathbf{x}_{sparse}\) with severe artifacts, such as streak artifacts, due to incomplete sinogram information. These artifacts degrade image quality and hinder accurate diagnosis.

%%%%%%%%%%%%%%%%%%%%%%%%%%%%%%%%%%%%%%%%%%%%%%%%%%%%%%%%%%%%%%%%%%%
%%%%%%%%%%%%%%%%%%%%%%%%%%%%%%%%%%%%%%%%%%%%%%%%%%%%%%%%%%%%%%%%%%%

Our goal is to reconstruct a high-quality full-view image \(\hat{\mathbf{x}}\) from degraded sparse-view data \(\mathbf{x}_{sparse}\), which can be formulated as a statistical inverse problem:
\[
\hat{\mathbf{x}} \sim p(\mathbf{x} \mid \mathbf{x}_{\text{sparse}}),
\]
where \(p(\mathbf{x} \mid \mathbf{x}_{\text{sparse}})\) is the posterior distribution derived using Bayes’ theorem:
\[
p(\mathbf{x} \mid \mathbf{x}_{\text{sparse}}) \propto p(\mathbf{x}_{\text{sparse}} \mid \mathbf{x})p(\mathbf{x}),
\]
with \(p(\mathbf{x})\) as the prior distribution representing the statistical characteristics of \(\mathbf{x}\), and \(p(\mathbf{x}_{\text{sparse}} \mid \mathbf{x})\) as the likelihood describing the degradation model. Sparse-view image \(\mathbf{x}_{\text{sparse}}\) often suffers from blurry and artifacts, necessitating advanced reconstruction algorithms to compensate for information loss and improve image quality.

\subsection{PFGM++ framework}
\begin{figure}
    \centering
    \includegraphics[width=0.5\linewidth]{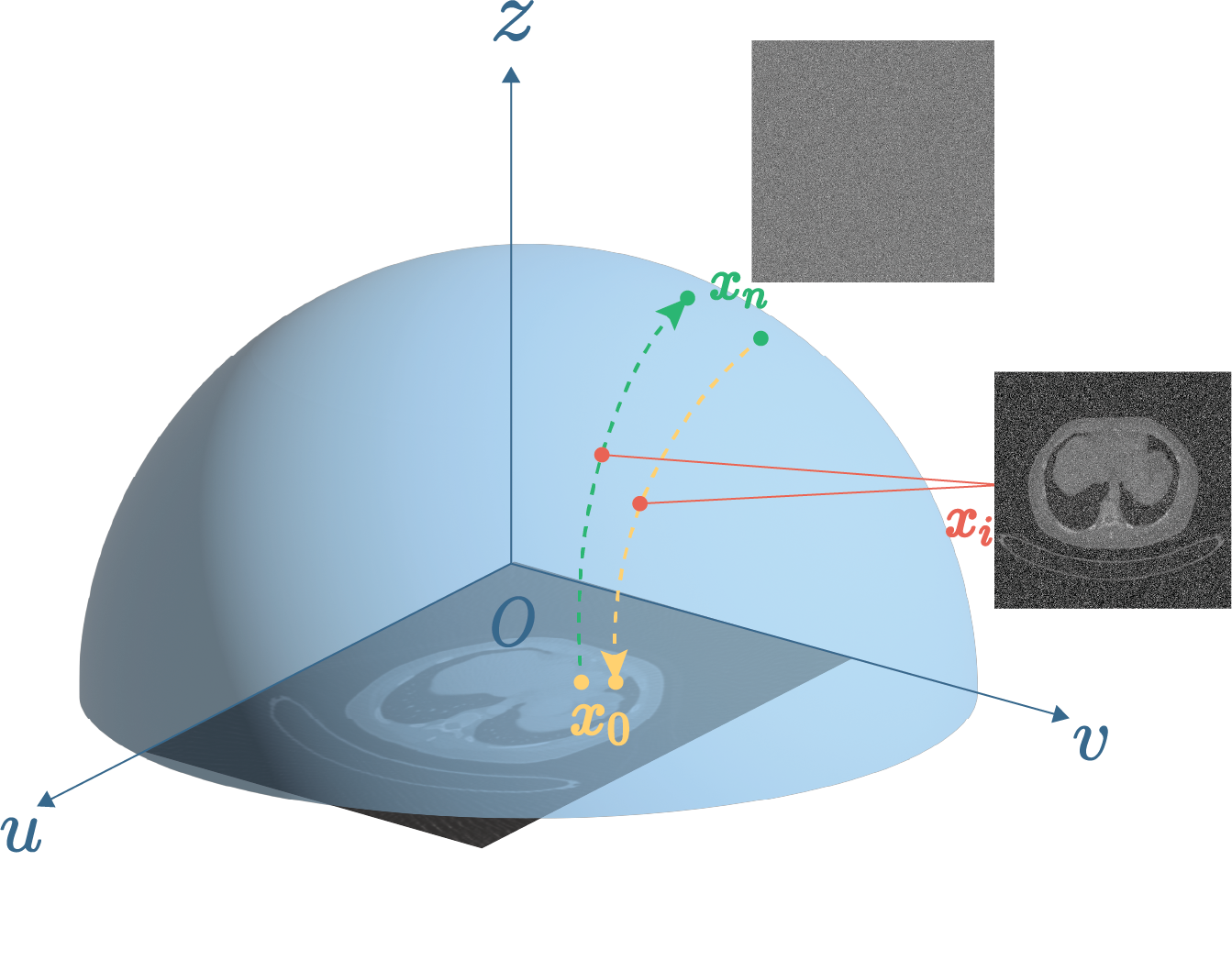}
    \caption{Illustration of 3D Poisson field trajectories for a 2D CT image distribution. The evolvements of a distribution or an augmented sample by the forward/backward ODEs pertains to the Poisson field.}
    \label{fig:pfgm}
\end{figure}
Instead of estimating a time-dependent score function like score-based diffusion models, PFGM++ focuses on the high-dimensional electric field, augmented the \textit{N}-dimensional data to \textit{N+D}-dimensional space (see the Fig. \ref{fig:pfgm} for a \textit{3D} Poisson field, here \(N=2, D=1\)):
\begin{equation}
\mathbf{E}(\tilde{u}) = \frac{1}{S_{N+D-1}(1)} \int \frac{\tilde{u} - \tilde{v}}{\|\tilde{u} - \tilde{v}\|^{N+D}} p(v) \, d\tilde{v},
\end{equation}
where \(p(v)\) is the ground truth data distribution, \(S_{N+D-1}(1)\) is the surface area of the unit \((N+D-1)\)-sphere, \(\tilde{v} := (v, 0) \in \mathbb{R}^{N+D}\), and \(\tilde{u} := (u, z) \in \mathbb{R}^{N+D}\) represent augmented data. In this framework, data points are treated as electric charges in augmented space, establishing a surjection between the ground truth data distribution and a uniform distribution on the infinite \((N+D)\)-dimensional hemisphere. The electric field's rotational symmetry on the \(D\)-dimensional cylinder \(\sum_{i=1}^{D} z_i^2 = r^2\) allows a reduction in dimensionality. By tracking \(r = r(\tilde{u})=\|\mathbf{z}\|_2\), the augmented variables are redefined as \(\tilde{v} := (v, 0) \in \mathbb{R}^{N+1}\) and \(\tilde{u} := (u, r) \in \mathbb{R}^{N+1}\). Here, the perturbation radius $r$ in high-dimensional space is a key parameter that guides data points from the initial noise distribution to the target distribution. Along the electric field line trajectories, the interest ODE is:
\begin{equation}
\frac{du}{dr} = \frac{\mathbf{E}(\tilde{u})_u}{\mathbf{E}(\tilde{u})_r},
\end{equation}
where, $\mathbf{E}(\tilde{u})_u = \frac{1}{S_{N+D-1}(1)} \int \frac{u - v}{\|\tilde{u} - \tilde{v}\|^{N+D}} p(v) \, dv, \quad$ $\mathbf{E}(\tilde{u})_r = \frac{1}{S_{N+D-1}(1)} \int \frac{r}{\|\tilde{u} - \tilde{v}\|^{N+D}} p(v) \, dv$.
This reduction establishes a bijection between data on the \(r = 0\) hyperplane (\(z = 0\)) and a distribution on the \(r = r_\text{max}\) hyper-cylinder. PFGM++ adopts a perturbation objective similar to the denoising score-matching objective in diffusion models. Using a perturbation kernel \(p_r(u|v)\), the objective is:
\begin{equation}
\label{equ:loss pfgm}
\mathbb{E}_{r \sim p(r)} \mathbb{E}_{v \sim p(v)} \mathbb{E}_{u \sim p_r(u|v)} \left\| f_\theta(\tilde{u}) - \frac{u - v}{r / \sqrt{D}} \right\|_2^2,
\end{equation}
where \(p(r)\) is the training distribution over \(r\). By choosing perturbation kernel \(p_r(u|v) \propto (||u - v||_2^2 + r^2)^{(N+D)/2}\), the minimizer of this objective aligns with the ODE is $f_\theta^*(\tilde{u}) = \sqrt{D} \, \frac{\mathbf{E}(\tilde{u})_u}{\mathbf{E}(\tilde{u})_r}.$ Starting from an initial sample from \(p_{r_\text{max}}\), samples from the target distribution can be generated by solving the ODE solver numerically.

\subsection{Posterior guidance Poisson flow}

In PFGM++ framework, the network aims to learn the mapping from a prior noise distribution to the target data distribution as shown in Eq.(\ref{equ:loss pfgm}). By optimizing this objective function, the network learns to generate samples from the target distribution. However, for the task of reconstructing high-quality full-view image $\hat{\mathbf{x}}$ from sparse-view image $\mathbf{x}_{spares}$, it is crucial to incorporate contextual information from sparse images. To achieve this, we introduce a conditional generation mechanism that utilizes the conditional image \( \mathbf{x}_{sparse} \) to guide the generation process. The network’s objective is extended to learn the mapping from the conditional distribution \( p(\mathbf{x}|\mathbf{x}_{sparse}) \) to the target distribution. Specifically, we extend the unconditional objective into the conditional form:
% \begin{equation}
% %\small
%  \mathcal{L}(\theta) = \mathbb{E}_{r, x, c, x_{perturb} \sim p_r(x_{perturb}|x)} \left[ \| f_\theta(x_{perturb}, r, x_{sparse}) 
%  - \frac{x_{perturb} - x}{r / \sqrt{D}} \|^2 \right] .  
% \end{equation}
\begin{multline}
\small
\mathcal{L}(\theta) = 
\mathbb{E}_{r, x, c, x_{\text{perturb}} \sim p_r(x_{\text{perturb}}|x)} \Bigg[ \\
\Big\| f_\theta(x_{\text{perturb}}, r, x_{\text{sparse}}) 
- \frac{x_{\text{perturb}} - x}{r / \sqrt{D}} \Big\|^2 
\Bigg].
\end{multline}

The key modification lies in introducing the conditional image \( \mathbf{x}_{sparse} \), making the network's output explicitly dependent on the perturbation sample \( \mathbf{x}_{perturb} \) (or the augmented data), the scale parameter \( r \), and the conditional input \( \mathbf{x}_{sparse} \). By incorporating conditional input \(\mathbf{x}_{sparse} \), the network can effectively leverage contextual information to guide the generation process and better address the inverse problem of sparse-to-full view image reconstruction. To facilitate efficient training, the objective is further discretized for batch optimization:
\begin{equation}
\mathcal{L}(\theta) = \sum_{i=1}^B \lambda(\sigma_i) \| D_\theta(\hat{x}_i, \sigma_i, \mathbf{x}_{sparse,i}) - x_i \|^2,
\end{equation}
where \( \hat{x}_i = x_i + R_i q_i \) represents the perturbed sample, \( R_i = \sigma_i \sqrt{D} \), and \( q_i = \frac{h_i}{\|h_i\|} \) is the normalized directional vector, $h_i$ is the random vector sampled in standard normal distribution $\mathcal{N}(0, I)$ for generating perturbation directions. \( \lambda(\sigma_i) \) is a noise-scale weighting function that balances the contribution of samples with different noise scales to the overall loss. And \( D_\theta(\hat{x}_i, \sigma_i,\mathbf{x}_{sparse,i}) \) is the network output that estimates the conditional score.

In the generation process,  a "hijack" strategy is used in \cite{hein2024ppfm} to generate image in a single step. While it greatly improves the generation efficiency, this method is for low-dose denoising, and it is infeasible to use the hijacking strategy on the sparse view images due to structural artifacts. Hence, we employ a stepwise generation strategy (full sampling), iteratively transferring the initial noise distribution to the target distribution. Stepwise generation retains the complete diffusion pathway and adapts the optimization techniques from EDM to enhance efficiency and stability. Specifically, we integrate the updated prior noise distribution, the hyperparameter mapping \( r = \sigma \sqrt{D} \), and EDM's noise scale setting \( \sigma(t) = t \). Using the variable transformation \( dr = d\sigma \sqrt{D} = dt \sqrt{D} \), the generation process is further optimized as:
\begin{equation}
dx = \frac{f_\theta^*(\tilde{x}, c)}{\sqrt{D}} dr = f_\theta^*(\tilde{x}, c) dt.
\end{equation}
Here,   \( \tilde{x} = (x, r) \) denotes the extended high-dimensional data space, \( f_\theta^* \) represents the conditional network's score estimate, \( r = \sigma \sqrt{D} \) is the high-dimensional perturbation scale,  and \( \sigma(t) = t \) is EDM's noise scale configuration to ensure smooth noise decay over time.

%%%%%%%%%%%%%%%%%%%%%%%%%%%%%%%%%%%%%%%%%%%%%%%%%%%%%%%%%%%%%%%%%%%%%%%%%%%%%%%%
\section{Experiments and Results}
\subsection{Dataset}
The Mayo Clinic dataset from the AAPM Low-Dose CT Grand Challenge \cite{aapm_low_dose_ct_grand_challenge} is used to simulate sparse-view CT data. This dataset contains 10 patients’ CT images reconstructed on a 512 × 512 grid with a slice thickness of 1 mm and a D30 kernel (medium). The physical size of each slice is $400\times 400$ $mm^{2}$, leading to a pixel size of about $0.781 \times 0.781$ $mm^2$. Using those image slices as phantoms, we numerically generate full-view sinograms via Siddon’s ray-driven algorithm \cite{siddon1985fast} with a typical fan-beam CT configuration (FOV radius about 300 mm , source-to-center: 550 mm, center-to-detector: 400 mm, detector interval  about 1/512 $rad$, 512 elements with equal-angle). Sparse-view sinograms are obtained by uniformly sampling 128 views from the full-view sinograms, and sparse-view images are reconstructed using filtered backprojection (FBP).  For training, we use 5,410 slices from 9 patients, reserving 526 slices from patient L506 for validation. This simulation framework ensures controlled evaluation of sparse-view CT reconstruction performance.

\subsection{Experimental settings}
The Poisson flow models were trained using the Adam optimizer following the EDM framework \cite{karras2022elucidating}. Training was conducted for 3 million iterations with a learning rate of \(2 \times 10^{-4}\), a batch size of 16, and random 256 × 256 pixel patches to reduce memory usage and provide data augmentation. Experiments were performed on a single NVIDIA RTX 6000 Ada GPU (48 GB). The DDPM++ architecture utilized a channel multiplier of 128, a channel distribution of [1, 1, 2, 2, 2, 2, 2], self-attention layers at resolutions 16, 8, and 4, along with preconditioning, exponential moving average (EMA), and 15\% augmentations. In PFGM++, the dimensional parameter \(D\) determines the data expansion into higher-dimensional space. This study focuses on \(D = 128\) and 16 sampling steps, selected via grid search over \(D \in \{64, 128, \infty\}\) and sampling steps \(\in \{8, 16, 32\}\). These settings provided the best balance between reconstruction quality and computational efficiency.

For evaluation, we used peak signal-noise ratio (PSNR), structural simularity (SSIM), and learned perceptual image patch similarity (LPIPS) with AlexNet. While PSNR and SSIM are standard in CT denoising, LPIPS aligns better with human perception, served as the primary metric for image similarity.

\subsection{Results}
To evaluate the effectiveness of our proposed PoCGM, experiments are conducted on sparse-view CT reconstruction using the AAPM challenge dataset with 128 projection views. Three advanced reconstruction methods, Red-CNN \cite{chen2017low}, SwinIR \cite{liang2021swinir}, and FBPConvNet \cite{jin2017deep}, are used as baselines due to their robust performance in noise suppression, artifact removal, and detail preservation.

Experimental results indicate that while Red-CNN and FBPConvNet effectively suppress artifacts, they struggle to restore fine details. SwinIR achieves better details but still retains streak artifacts and misses some structures under extremely sparse conditions. In contrast, PoCGM consistently outperforms all baselines, demonstrating superior artifact suppression and detail preservation in different settings.

\begin{figure}
    \centering
    \includegraphics[width=1\linewidth]{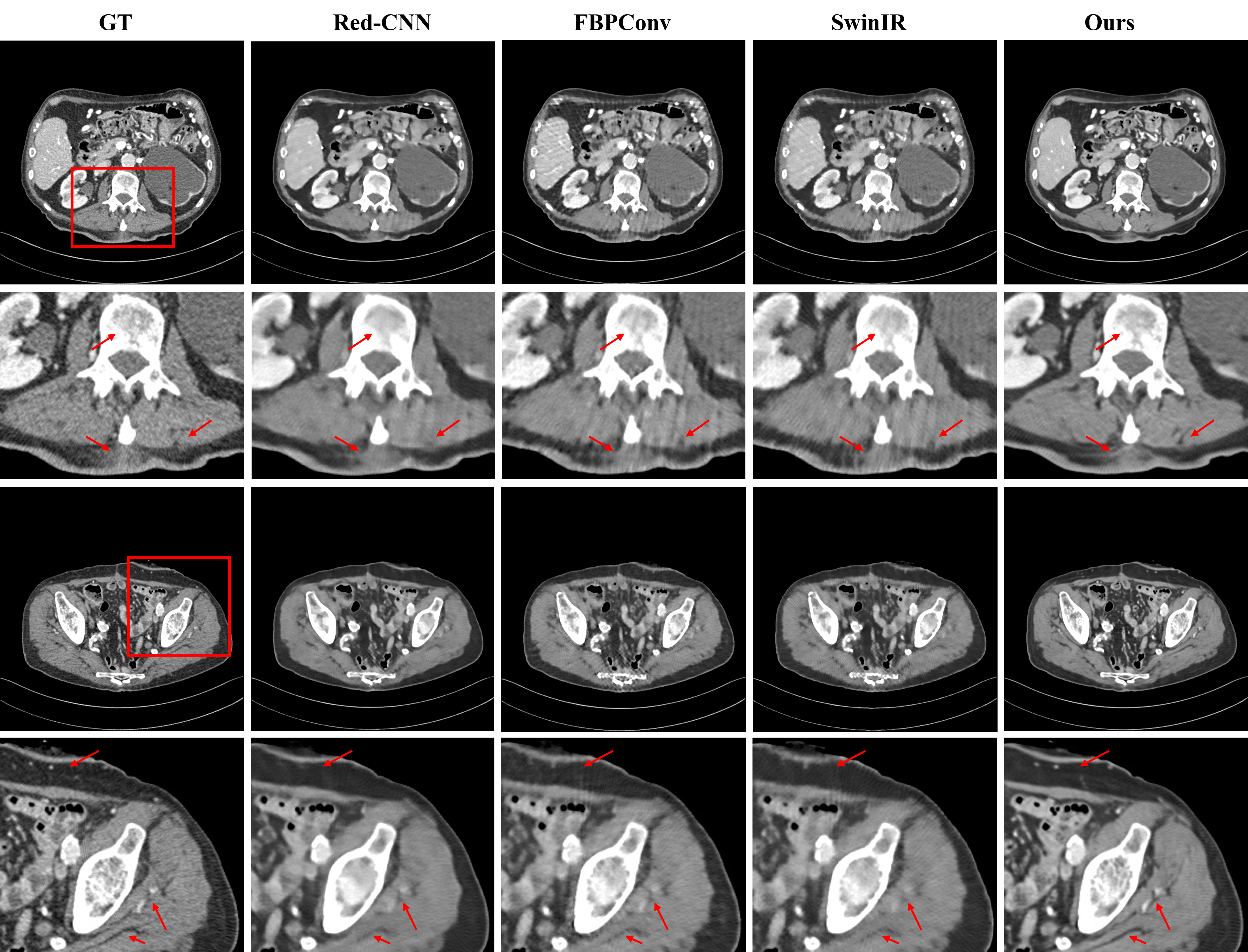}
    \caption{Representative reconstruction results from 128 views using different methods. The $2^{nd}$ and $4^{th}$ row are the ROIs of the $1^{st}$ and $3^{rd}$ rows. The display window is [-160,240] HU.}
    \label{fig:Comparison}
\end{figure}

Table \ref{table: comparison} shows PoCGM’s significant improvements in PSNR, SSIM, and LPIPS, while Fig. \ref{fig:Comparison} illustrates its ability to reconstruct images with richer details and fewer artifacts. Unlike SwinIR, which focuses on local features, PoCGM benefits from its Poisson-flow-based generative prior, leveraging supervised learning to generate a high-quality conditional prior from the sparse-view projection data and refining the reconstruction through iterative consistency optimization in the image domain. This unified approach ensures high-quality reconstructions by effectively balancing noise reduction, artifact suppression, and texture preservation, while maintaining fidelity to sparse-view observations.

%%%%%%%%%%%%%%%%%%%%%%%%%%%%%%%%%%%%%%%%%%%%%%%%%%%%%%%%%%%%%%%%%%%%%%%%%%

%%%%%%%%%%%%%%%%%%%%%%%%%%%%%%%%%%%%%%%%%%%%%%%%%%%%%%%%%%%%%%%%%%%%%%%%%%
%%%%%%%%%%%%%%%%%%%%%%%%%%%%%%%%%%%%%%%%%%%%%%%%%%%%%%%%%%%%%%%%%%%%%%%%%%
\begin{table}
\centering
\caption{Reconstruction quantitative indicator of AAPM validation set at 128views. ↓ Means lower is better. ↑ Means higher is better. Best results in bold. }
\resizebox{\columnwidth}{!}
{
\begin{tblr}{
  cells = {c},
  hline{1} = {-}{0.08em},
  hline{2} = {-}{0.05em},
  hline{5} = {-}{},
}
                               & Red-CNN & FBPConvNet & SwinIR  & PoCGM(Ours) \\
PSNR($\uparrow$) & 43.93~ & 40.45~     & 42.30   & \textbf{45.64}  \\
SSIM($\uparrow$)                           & 0.969~ & 0.966      & ~0.955~ & \textbf{0.979}  \\
LPIPS($\downarrow$)                          &       0.0350 &     0.0352~       & 0.0476 & \textbf{0.0345} 
\end{tblr}
}
\label{table: comparison}

\end{table}

%%%%%%%%%%%%%%%%%%%%%%%%%%%%%%%%%%%%%%%%%%%%%%%%%%%%%%%%%%%%%%%%%%%%%%%%%%

\section{Discussion and Conclusion}
This study introduces PFGM++ for sparse-view CT reconstruction, leveraging a conditional generation–based posterior sampling strategy to suppress streak artifacts and noise from sparse-view projections. Unlike diffusion models like DDPM, which require thousands of sampling steps, PFGM++ employs a deterministic ODE sampling process, enabling high-quality reconstructions with only 16 sampling steps. This highlights its inherent advantage in sampling efficiency, making it suitable for dose-sensitive and time-critical imaging scenarios. Additionally, the physics-inspired Poisson flow framework enhances detail preservation and artifact suppression, demonstrating its applicability to challenging CT reconstruction tasks.

The posterior sampling strategy flexibly integrates reconstructed outputs with real projection data, balancing noise suppression with texture fidelity. While larger ODE step sizes may introduce local errors, PFGM++ is less sensitive to step size variations, enabling effective reconstruction even with aggressive hyperparameter settings. Although sampling time is not explicitly measured, the reduced step count underscores its computational efficiency compared to diffusion models.

By integrating posterior sampling into PFGM++, this work provides an efficient and robust solution for sparse-view CT reconstruction, excelling in artifact suppression, detail retention, and practical adaptability.

\bigskip
\textbf{\color{black} Acknowledgment:} The work was supported in part by NIH/NIBIB grants R01EB034737 and R01EB032807.

%-------------------------------------------------------------------------------------------
\printbibliography
\end{document}